# THACO, a Test Facility for Characterizing the Noise Performance of Active Antenna Arrays

E. E. M. Woestenburg, L. Bakker, M. Ruiter, M. V. Ivashina, and R. H. Witvers

ASTRON, The Netherlands Institute for Radio Astronomy, The Netherlands

woestenburg@astron.nl

*Abstract*—This paper discusses an outdoor test facility for the noise characterization of active antenna arrays, using measurement results of array noise temperatures in the order of 50 K for a number of small aperture arrays. The measurement results are obtained by a Y-factor method with hot and a cold noise sources, with an absorber at room temperature as the hot load and the cold sky as the cold load. The effect of shielding the arrays by the test facility, with respect to noise and RFI from the environment, will also be discussed.

## I. INTRODUCTION

THE next generation telescope for radio astronomy, the Square Kilometre Array (SKA), will have unprecedented sensitivity, resolution and survey speed, with an envisioned collecting area of one square kilometer ([1], [2]). It will consist of a large number of receiving stations with approximately 100 m diameter, filled with conventional reflector telescopes with 10-15 m diameter, as well as large fields of flat antennas, establishing phased array stations. The latter, the Aperture Array (AA, [3], [4]) concept for the SKA mid-frequency range from 400 MHz to 1500 MHz will use up to 10000 small antennas and LNAs in a station, arranged in tiles of 1 m². Another concept will use Phased Array Feeds (PAFs, [5]), consisting of a 1 m² tile with hundreds of antennas and LNAs in the focus of the large reflectors. For both array systems room temperature LNAs will be closely integrated with the array antenna elements, without the possibility to measure the individual properties of the antenna array and LNAs, because they cannot be separated physically and are directly matched to each other without intermediate matching to the 50 Ω standard impedance. For the purpose of this paper, we will define the combination of such an antenna array, 'integrated' with LNAs, as an 'Active Antenna Array'. The advantage of such arrays is that they potentially have larger bandwidth and sensitivity, because simple, low loss matching circuits may be used in the design of the antenna-LNA interface. For the measurement of the noise properties of these active arrays an alternative must be used for the standard hot/cold Y-factor method ([6], [7]), generally using room temperature and cryogenic coaxial or waveguide loads, or noise measurement systems with a solid state noise source. The alternative method described in [6] uses a room temperature absorber covering the array aperture as a hot load and the sky as a cold load. Depending on the beam width of the antenna pattern, during the cold load measurement, noise from the environment will be coupled to the array, as well as interfering RF-signals from mobile communication, radar transmitters and broadcasting stations, which may deteriorate the quality of the noise measurement. Therefore a shielding funnel as shown in Fig. 1 was designed and constructed to reduce and possibly eliminate these effects on the noise measurements. This THot And COld (THACO) test facility has been introduced in [8], which also shows the result of a preliminary measurement on a PAF prototype element. This paper presents an assessment of the usefulness of THACO for the characterization of the noise temperature of low noise array systems (of the order 50 K noise temperature) with different numbers of active elements, in the 1000 MHz to 2000 MHz frequency range. For this purpose it uses noise measurements on a single element and on active arrays with 4- and 16-elements, located inside and outside THACO. The shielding of THACO for RF-signals will also be discussed.

## II. MEASUREMENT METHOD AND PROCEDURE

The principle of the noise measurement is the hot/cold Y-factor method, which is the origin of the name THACO. THACO consists of a metal funnel with removable roof, covered with RF absorbing material for the 'hot load' measurements. The roof can be moved away from the funnel on a short rail-track, to expose the array to the sky for the 'cold load' measurement. Fig. 1a shows THACO with the roof in the 'hot load' position. Fig. 1b shows the position of the roof for the 'cold load' measurement.

For the noise measurements a prototype array tile was used with a total of 144 Vivaldi antenna elements and LNAs, configured in a two-dimensional array with 8x9 elements with 11 cm spacing, for each of the two linear polarizations ([9]). Details of this system for use as a PAF can be found in [5]. For the noise measurements to characterize THACO the tile was used as an aperture array and was placed on the ground inside with the roof in the 'on' and 'off' positions. The measurement outside THACO was done at the location of the roof in the 'off' position (see Fig. 1b) for the 'hot' load. The 'cold' load measurement outside THACO was done with the roof in the position of Fig. 1a.

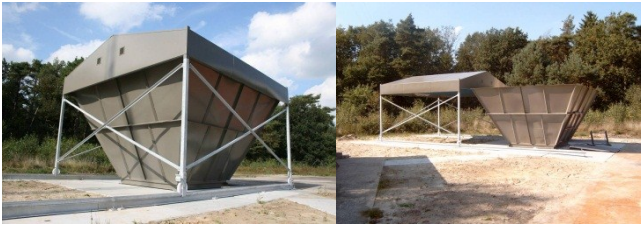

**Fig. 1  a)  b) THACO, a) with roof in the 'hot load' position and b) with roof removed, for the 'cold load' measurement**

To obtain measurement results two small arrays were formed using analog beam forming with only 2x2 and 4x4 elements in the centre of the tile. The output signals from the LNAs of the 2x2 and the 4x4 array were combined with a proper arrangement of in-phase combiners, forming broadside beams. The analog output signals of the beam formers were fed to the receiver of an Agilent Noise Figure Meter 8970B and the noise temperature was determined, using the hot and cold loads as illustrated in Fig. 1. The noise output powers measured by the receiver as a function of frequency were stored and used to calculate the Y-factor and subsequently the array noise temperature.

Two other types of measurements have been done to characterize the THACO test facility. The first determines the influence of the metal funnel on the reflection coefficient of an antenna placed inside it. This was done to see if the metal housing of THACO would change the reflection coefficient of the antenna, because possible antenna impedance changes at the LNA input may influence the LNA noise temperature. Three different types of antennas have been measured, observing the change in reflection coefficient with a Network Analyzer as the antenna was placed inside THACO.

Shielding the array from interfering signals is important to reduce the influence of strong RFI-signals on the noise measurement. The result will also give an indication of possible suppression by THACO of noise from the environment, which couples to the array system. Measurements have been done with the roof of THACO open and closed in two different measurement setups, as described in detail in section III.B.

### III. MEASUREMENT RESULTS

*A. Antenna reflection coefficient*

The three antennas measured were a monopole with a 14 db return loss at 1025 MHz, a single Vivaldi antenna element (the same as in the array for the noise measurements) with a 10 dB return loss at 1400 MHz and a horn antenna with better than 13 dB return loss over the 1000 MHz to 2000 MHz band. All antennas have been measured over a frequency range between 500 MHz and 2000 MHz and show no significant change in reflection coefficient at different locations inside THACO, compared to a location outside. Because there are no significant differences due to THACO, the measurement result is displayed in Fig. 2 only for the Vivaldi antenna. The conclusion is that the noise measurements will not be influenced by changes in the antenna impedance by placing the array inside THACO.

*B. RFI-suppression*

Two measurement setups were used to determine the RFI-suppression by THACO. The first one used the 8 MHz

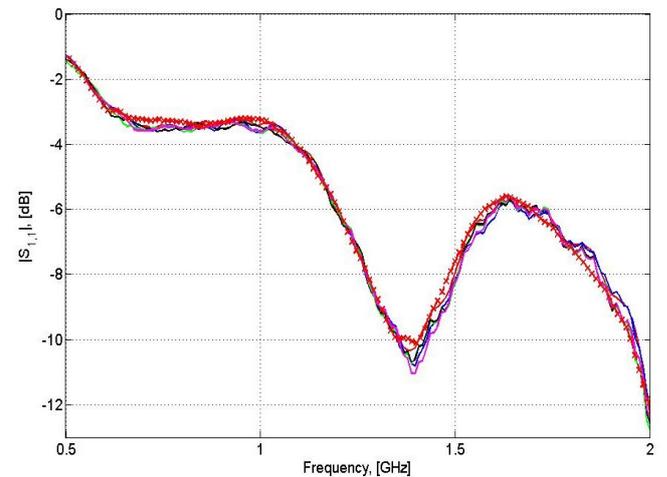

**Fig. 2 Measured reflection coefficient (return loss) for the Vivaldi antenna, inside and outside THACO**

DVB-S signal at 800 MHz from a local TV-station, which was detected with a printed circuit board Vivaldi antenna. The maximum power outside THACO, measured with this antenna and a spectrum analyzer, was -75 dBm. With the roof closed a maximum of -99 dBm was measured inside, giving a suppression of 24 dB. With the roof removed the level inside was only 5 dB lower than outside, showing limited shielding by THACO for a source on the horizon.

The second measurement was done using a transmitter with a broad band log per antenna at 60 m distance from THACO, producing a sweeping signal between 1000 MHz and 2000 MHz. The same Vivaldi antenna was used for detection of the signal and the measurements confirmed the limited shielding of the test facility, which gave on average a suppression of 10 dB with the roof closed. With the roof open the suppression varies between 0-5 dB as a function of frequency. This means that, in case of strong RFI, the 'hot' noise measurement may be influenced by the RFI-signal. The conclusion based on these results is that there is limited shielding from RFI by THACO for the presented measurement setups, with a single Vivaldi receiving antenna, which has a broad beam which apparently interacts with the walls of THACO and couples to the RFI-environment. For antennas with narrower beams it is expected that this effect will be reduced. This is subject of further studies and requires accurate characterization of the antenna patterns in the presence of the shielding test facility.

The limited suppression by THACO of RFI-signals and the coupling to the antenna inside is also an indication for the coupling of (approximately 300 K) noise from the environment, which may have a large impact on the noise measurement results for antennas with broad beam patterns.

*C. Noise measurements*

An important argument for the realization of THACO has been the expected suppression of noise from the environment (ground, trees). Simulations in [10] give an estimate of these noise contributions. Because of the computational burden in modeling a large structure such as THACO, its scattering properties (due to the antenna as a source positioned inside it) have been analyzed based on the simulations at frequencies below 500 MHz. Using the outcome of these simulations THACO was designed, but measurements in the 1000-2000 MHz frequency band must reveal whether or not it will meet its goal of adequate noise suppression for this frequency range.

To assess THACO properties hot/cold noise measurements have been done between 1000 MHz and 2000 MHz, with a single (stand-alone) antenna element, one embedded element in the centre of the array and 4-element and 16-element arrays, also in the centre of the tile.

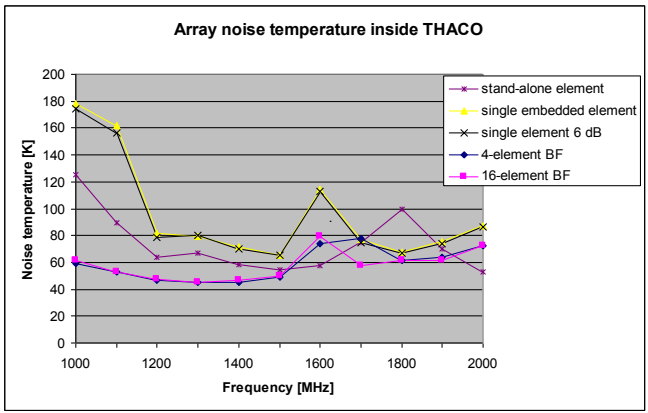

**Fig. 3 Measured noise temperatures inside THACO for single elements and two arrays**

Measurements were done with the antennas inside and outside THACO for beams directed at broadside. Figs. 3 and 4 show the results from the measurements, respectively inside and outside THACO. Fig. 3 also shows the result for a single element with a 6 dB attenuator between the LNA output and the input of the measurement system, to illustrate the effect of the receiver noise contribution and possible non-linearity. The results are equal and show no sign of a second stage contribution or non-linearity. Fig. 4 includes a result for the 4-element array, rotated with 90° with respect to its vertical axis, which shows that the position of the array inside THACO hardly has influence on the results.

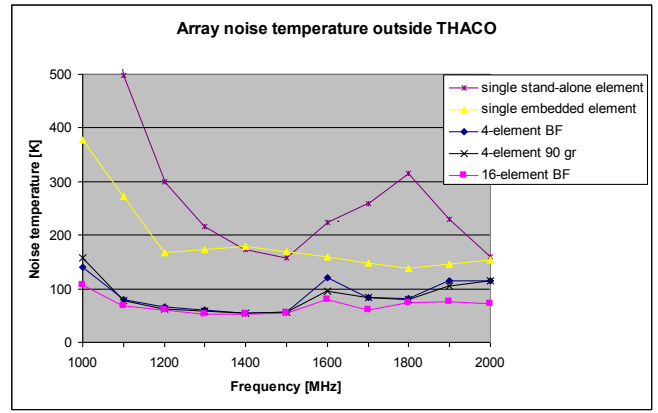

**Figure 4 Noise temperatures outside THACO for the same antennas as in Fig. 3**

The noise temperatures for the 4- and 16-element arrays inside THACO are almost equal (Fig. 3), contrary to outside (Fig. 4) where the noise for the 16-element array is lower. The single elements show much higher noise temperatures, both inside and outside THACO. Fig. 5 compares the results from measurements for the 4-element and 16-element arrays in- and outside, at the same temperature scale. From Figs. 3-5 the shielding effect of THACO is clearly visible for the single elements and the 4-element array, but it also shows a limited suppression of the noise from the environment for the single elements. Even in the measurement of the 16-element array THACO gives some suppression at low frequencies. Fig. 5 also compares the measurements with a simulated result for the full array, which is in close agreement with the measured results inside THACO between 1200 MHz and 1600 MHz.

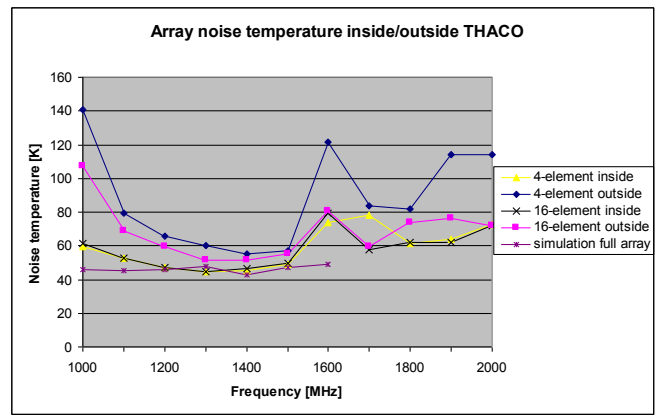

**Fig. 5 Comparison of measurement results in- and outside THACO**

Fig. 6 compares the measured noise temperatures for a single embedded element with the simulation for that element and with the result for a single stand-alone element, all inside THACO. The simulation is in good agreement with the measurement as a function of frequency, but shows some differences in absolute values. This is because the numerical model neglects the effects of the trees and the buildings in the description of the noise temperature due to external noise sources.

## IV. DISCUSSION OF RESULTS

The suppression by THACO of noise pick up from the environment appears most effective for the single elements and the small 2x2 array, but is not sufficient to fully eliminate noise pick up for the single elements, as one would then expect similar noise performance for a single element and the (elements of the) arrays. This conclusion is consistent with the limited RFI-suppression, discussed in section III.B, which indicated non-ideal suppression of ambient noise. For the 4- and 16-element arrays the noise measurement results inside THACO seem to be a correct representation of the noise properties of the arrays. Except for frequencies below 1200 MHz, the use of THACO does not result in better measured noise properties for the 16-element array.

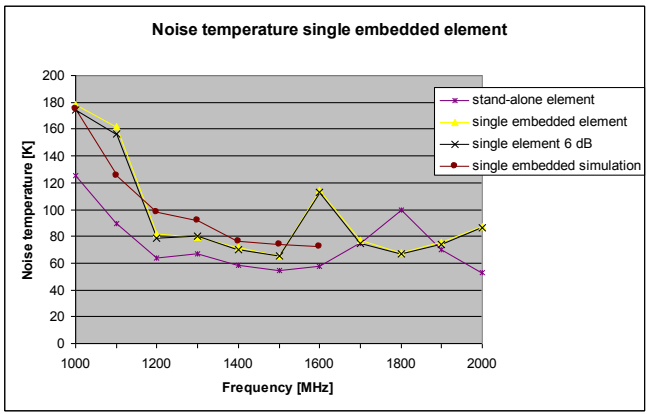

Fig. 6 Comparison of measurement results for single elements inside THACO with the simulation result for a single embedded element

## V. CONCLUSION

The presented measurement results and their discussion lead to the conclusion that the use of THACO for this type of measurements is meaningful for the measurement of small arrays and single antenna elements with broad beams. For larger arrays (4x4 and larger) measurement results are comparable and consistent with the simulations which neglect the effect of an external noise contribution due to the trees. For those cases the use of THACO is not necessary to obtain reliable noise measurements. On the other hand THACO is a user friendly facility with a removable absorbing roof on a rail- track and it does not introduce negative effects on the measurements. It does suppress the effects of noise pick up from the environment, however not to the level that these effects are negligible and that the measurement of a single antenna is a good representation of the noise behavior of an array. However, the results for small 2x2 and 4x4 arrays inside THACO do provide a reliable prediction of the performance of a larger array. These conclusions are based on noise measurements in the 1000 MHz to 2000 MHz frequency range. Measurements at lower frequencies have not yet been done, but the results towards 1000 MHz show a tendency to a larger influence of the environment at frequencies below 1000 MHz.

ACKNOWLEDGMENT

This work has benefited from funding by the European Community's sixth Framework Program.